\documentclass[sigconf,nonacm]{acmart}
\AtBeginDocument{%
  \providecommand\BibTeX{{%
    \normalfont B\kern-0.5em{\scshape i\kern-0.25em b}\kern-0.8em\TeX}}}

\newtheorem{mydef}{Definition}

\usepackage[ruled,vlined, linesnumbered]{algorithm2e}
\usepackage{siunitx}
\usepackage{booktabs}
\usepackage{subfig}

\begin{document}

\title{Solving the Online Assignment Problem with Machine Learned Advice}


\author{Clarence Gabriel R. Kasilag}
\email{crkasilag@up.edu.ph} 
\affiliation{%
  \department[0]{Algorithms and Complexity Laboratory}
  \department[1]{Department of Computer Science}
  \institution{University of the Philippines Diliman}
     \country{Philippines}
}

\author{Pollux M. Rey}
\email{pmrey@up.edu.ph}
\affiliation{%
  \department[0]{Algorithms and Complexity Laboratory}
  \department[1]{Department of Computer Science}
  \institution{University of the Philippines Diliman}
   \country{Philippines}
}
\author{Jhoirene B. Clemente}
\email{jbclemente@up.edu.ph}
\affiliation{%
  \department[0]{Algorithms and Complexity Laboratory}
  \department[1]{Department of Computer Science}
  \institution{University of the Philippines Diliman}
     \country{Philippines}
}
\renewcommand{\shortauthors}{Clemente, Kasilag and Rey}

\begin{abstract}
The online assignment problem deals with assigning n elements of one set to another set which arrives sequentially on a one-to-one basis. This problem is also known as the online weighted bipartite matching which produces the smallest weight perfect matching possible. It plays an important role in the fields of operational research and computer science which is why immense attention has been given to improve its solution quality. Due to the incomplete information about the input, it is difficult for online algorithms to produce the optimal solution. The quality of the solution of an online algorithm is measured using a competitive ratio. It has been proven that for this problem, no online deterministic algorithm can achieve a competitive ratio better than (2n-1). It has been shown that advice in online computation improves the lower bound of the competitive ratio of online problems. Advice in online computation can be interpreted as additional information for the online algorithm to compensate for the lack of information about the whole input sequence. In this study, we investigate how introducing machine-learned advice could improve the competitive ratio for this problem. We provide an online algorithm for the online assignment problem by simulating a machine learning algorithm that predicts the whole input in advance. We utilize an optimal offline algorithm to provide a matching solution from the predicted input. Furthermore, we investigate how the prediction error of machine learning affects the competitive ratio of the online algorithm. We utilize a benchmark data set to perform our empirical analysis. We show that as the Machine Learning prediction error increases, the solution quality decreases. Moreover, the magnitude of error is directly proportional to the size of the input. This result is analogous to the competitive ratio of the best deterministic algorithm for the online assignment problem which is dependent also on the parameter n.
\end{abstract}

\ccsdesc[500]{Theory of computation~Online algorithms}
\ccsdesc[500]{Computing methodologies~Machine learning}
\ccsdesc[100]{Applied computing~Operations research}

\keywords{online assignment problem, machine-learned advice, competitive analysis}


\maketitle

\section{Introduction}
    The \textit{assignment problem} has diverse applications in various fields. It is used by industries to assign jobs to workers, by transportation companies in assigning passengers to vehicles, and among others. The problem works by having two disjoint sets of nodes with every node from one set connected to a node from the the other set by an edge such that the total weight of the edges is optimal (either minimum or maximum). The most widely used offline algorithm to solve the problem is the \textit{Hungarian Algorithm} which runs at $\text{O} (n^4)$ time, however, there are better algorithms for the problem, in terms of time complexity, such as in \cite{karp1980algorithm} which runs at $\text{O} (n^2 \log n)$ time.
    
    We  focus on the online variant of this problem wherein the nodes from one set are given in advance, while the nodes from the other set arrive one at a time. This was introduced in \cite{khuller1994line}. This variant of the assignment problem imitates real-world situations as data arrives with respect to time. Online algorithms for this variant must decide on what to do with the arriving nodes immediately and the decisions made are irrevocable.
    
    The problem, however, is that online algorithms perform worse than their offline counterparts because of the lack of knowledge of the entire input sequence which leads to a less optimal solution. The \textit{competitive ratio} of an online algorithm is used as a metric to compare the online algorithm to the optimal offline algorithm for the problem. For the assignment problem, the best deterministic algorithm was from \cite{khuller1994line} and \cite{kalyanasundaram1993online}. It has a competitive ratio of  $(2n-1)$ and is proven to be the tight lower bound for all online deterministic algorithms for the problem. In terms of expectation, both $\text{O} (\log^2 n)$ and $\text{O} (\log^3 n)$-competitive randomized algorithm also exist for the problem \cite{bansal2007log} \cite{meyerson2006randomized}.
    
    In this paper, we aim to investigate if we could push these deterministic and randomized tight bounds ever further by incorporating an approach that uses Advice from Machine Learning. We aim to discover if an online algorithm using advice will perform better than the $2n-1$ benchmark for deterministic approaches and the $log(n)$ for randomized approaches, what its trade-offs are, and to analyze the results of our experimentation. 
    
    Now the question is, in advice, how would the \textit{oracle} provide information to our online algorithm? We harness the predictive properties of machine learning \cite{lykouris2018competitive}. By feeding data into a machine learning model, we can estimate the input up to a point that the difference between the actual online input and the generated input will be small enough for an offline algorithm to get a solution comparable to that of those obtained from optimal randomized and deterministic algorithms. 
    
    Throughout the paper, we will use the terms \textit{node} and \textit{vertex}; \textit{ML}, \textit{Machine Learned}, and \textit{Machine Learning}; and, \textit{input} and \textit{request} interchangeably.
\section{Preliminaries}
    Our study revolves around the Online Assignment Problem or Online Weighted Bipartite Matching, it is important at this point to define which variant we aim to investigate. 
    \begin{mydef}[Online Assignment Problem]\label{def:assignment}
        Given a complete bipartite graph $G = (U, V, E)$, where $U$ and $V$ are two disjoint sets of $n$ vertices, and $E = U \times V$. Assume that each edge in $E$ is associated with non-negative weights.
        
        Initially vertices in $U$ are known and vertices in $V$ arrive one at a time revealing its edge weights.
        
        The Online Assignment Problem is defined as follows: obtain a minimum weight perfect matching in an edge-weighted bipartite graph such that the following constraints are satisfied:
        
        \begin{enumerate}
            \item Real-time constraint: once a vertex in V arrives, a vertex in U must be immediately assigned to it before the next vertex in V arrives.
            \item Invariable constraint: once a vertex in U is assigned to a vertex in V, the assignment cannot be revoked.
        \end{enumerate} 
    \end{mydef}
    With this given, we now define the metric in which we measure the \textit{goodness} of an algorithm for our problem. This case, we have chosen to use the \textit{competitive ratio} of an algorithm as it measures the solution quality of an online algorithm and how it perform against a known optimal algorithm. 
    \begin{mydef}[Competitive Ratio]
        For all finite request sequences $I$, we define $\textsc{Alg}(I)$ to be the performance of an online algorithm $\textsc{Alg}$ and $\textsc{Opt}(I)$ similarly to be the performance of an offline algorithm $\textsc{Opt}$. $\textsc{Alg}$ has a competitive ratio of $c$ (or is $c$-competitive) if there exists a constant $b$ such that $$ \textsc{Alg}(I) \leq c \cdot \textsc{Opt}(I) + b $$
        
        If $b = 0$, $\textsc{Alg}$ has a strictly competitive ratio of $c$ (or is strictly $c$-competitive) such that $$ \textsc{Alg}(I) \leq c \cdot \textsc{Opt}(I) $$
    \end{mydef}
    \subsection{Related Work}
    
    The pursuit for a faster and more efficient algorithm has always been an interest in the computer science research space. This is not different from what we want to investigate, that is, to formulate an algorithm which results in a better solution quality for the online assignment problem. We do this by attempting to emulate some of the processes introduced by Lykouris and Vassilvitskii's model which merged online algorithms with machine-learned advice. 
    \\ \\
    \noindent \textbf{Optimal Offline Algorithm}. Our algorithm will involve using an optimal algorithm for solving the assignment problem. In such, various algorithms have been developed to efficiently solve the problem which is summarized in Tables 1 and 2. The \textit{Hungarian Algorithm} is one of the best known-algorithm to solve this classical problem. \cite{kuhn1955hungarian} presented the algorithm which was refined by \cite{munkres1957algorithms}. It was the first algorithm to solve the assignment problem in polynomial time, specifically at $O(n^4)$. 
    Further studies at this problem has resulted to producing an algorithm at time $O(n^3)$ in \cite{dinic1969algorithm}, \cite{tomizawa1971some}, and \cite{edmonds1972theoretical}. 
    
    In 1980, an $O(mn \log n)$ algorithm to solve the assignment problem for $m$ sources and $n$ destinations was discovered by \cite{karp1980algorithm}. Its tiome complexity was achieved under the assumptions that costs of the edges are independent random variables, and, the costs of the edges are connected to a source are drawn independently from a common distribution. In our case, in which we expect the sizes of $m$ and $n$ to be equal, the algorithm would have a running time of $O(n^2\log n)$
    
    \noindent \textbf{Online Algorithms}. In our study, we will be using known online algorithms as a benchmark on which we compare our algorithm in our empirical analysis. The first known online version of an edge-weighted bipartite matching problem was introduced independently by \cite{kalyanasundaram1993online} and \cite{khuller1994line}. 
    
    In this version, assuming that the bipartite graph is complete, a set of vertices, called \textit{girl vertices} are given in advance, while the other set, called \textit{boy vertices} arrive one at a time. When a boy vertex arrive, he reveals the weights of edges connected to him and the girl vertices, and, he has to be matched off immediately, this decision is irrevocable. Similar to the Linear Sum Assignment Problem, the goal of this algorithm is to minimize the sum of the obtained weights. 
    
    Both papers, \cite{kalyanasundaram1993online} and \cite{khuller1994line}, gave a $(2n-1)$-competitive online algorithm to solve the problem and proved that no online deterministic algorithm can achieve a competitive ratio lower than that for all metric spaces. 
    
    With the tightness of the competitive ratios proven, different approaches to achieving an optimal solution for the problem was done. One of these are into using \textit{Randomization} as suggested by the open problem mentioned in \cite{kalyanasundaram1993online}. 
    
    Furthermore, \cite{meyerson2006randomized} discovered an online randomized algorithm with an expected competitive ratio of $O\log^3n$-the first of this kind to achieve a poly-logarithmic ratio for the problem on general metrics. A year later, an online randomized algorithm with an expected competitive ratio of $O(\log^2 n)$ was discovered by \cite{bansal2007log} which improved upon the conversion from tree metrics to general metrics. 
    
\begin{table}[h]
  \caption{Summary of the some of the offline algorithms for AP}
  \label{tab:freq}
  \begin{tabular}{ccl}
    \toprule
    Algorithm&Time Complexity\\
    \midrule
    Kuhn, 1955 \cite{kuhn1955hungarian} & $O(n^4)$\\
    Munkres, 1957 \cite{munkres1957algorithms} & $O(n^4)$\\
    Tomizawa, 1971 \cite{tomizawa1971some} &$O(n^3)$\\
    Karp, 1980 \cite{karp1980algorithm} & $O(n^2 \log{n})$ \\
    Edmonds and Karp, 1972 \cite{edmonds1972theoretical} &$O(n^3)$ \\
  \bottomrule
\end{tabular}
\end{table}
\begin{table}[h]
  \caption{Some of the offline algorithms for AP}
  \label{tab:freq}
  \begin{tabular}{ccl}
    \toprule
    Algorithm&Time Complexity\\
    \midrule
    Khuller et al., 1994 \cite{khuller1994line} & $2^n - 1$\\
    Khuller et al., 1994 \cite{khuller1994line} & $2n - 1$\\
    Meyerson et al., 2006 \cite{meyerson2006randomized} & $\log^3 n$\\
    Bansal et al, 2007 \cite{bansal2007log} & $\log^2 n$ \\
  \bottomrule
\end{tabular}
\end{table}

    In this study, we would be using these deterministic and randomized algorithms and their resulting competitive ratios for the assignment problem as a benchmark for analyzing our algorithm. We would try to determine the trade-offs between using an algorithm with advice to the known deterministic and randomized algorithm in terms of solution quality, time complexity and overall efficiency and optimality of the algorithm. 
    
    \noindent \textbf{Machine Learned Advice}. The use of advice has been a relatively new technique in solving online algorithms. The main reasoning why such technique is used is because they provide sufficient knowledge to an online algorithm to have better responses to requests arriving sequentially such as those mentioned in \cite{dobrev2008much} (Oracle with answerer and helper modes), \cite{bockenhauer2009advice} (Advice Tape), and \cite{steffen2014advice} (Clairvoyant oracle with unlimited computational power). 
    
    We decided that for this problem, the most viable way on implementation is to use Machine Learning as a source of data to be used for the online algorithm. In this sense, \cite{lykouris2018competitive} conceptualized a framework on how to utilize this machine learning models in order to improve the performance of online algorithm. This perspective of using Machine Learned Advice to augment online algorithms have been applied to multiple problems. \cite{indyk2020online} used ML advice in optimizing the online page migration problem and discovered that the competitive ratio approaches 1 as the error rate diminish to 0. \cite{lykouris2018competitive}and \cite{rohatgi2020near} both tackled the caching problem, \cite{lattanzi2020online} and \cite{purohit2018improving} with the ski-rental problem, all of which resulted in an improvement to the competitive ratios of their respective problems. In this paper, we apply the same methodology with the assignment problem with hopes of investigating whether advice can find a solution better than the bounded deterministic competitive ratio of $(2n-1)$.
    
\section{Online Algorithm for the Assignment Problem with ML Advice}
    The thought process into conceptualizing the algorithm is that what if we have a certain oracle that could provide a \textit{good enough input} for an optimal offline algorithm for the problem, then certainly, a \textit{good enough} solution can be obtained. The following question has led to an algorithm that follows. 
    
    \begin{algorithm}
        \SetKwInOut{Input}{input}\SetKwInOut{Output}{output}
        \SetKwComment{Comment}{$\triangleright$\ }{}
        \SetAlgoLined
        \Input {Actual input $A$, an $nxn$ matrix, where $a_{ij} \in \mathbb{N}^{[1,n]}$}
        \Output{Matching $M$}
        \BlankLine
        \For{$i \gets 1$ \textbf{to} $n$} {
            $a'_i \gets$ $mlModel(a_{i-1}, A)$ \Comment*[r]{returns predictions subject to error}
        }
        $P = Karp(A')$, where $P = p_1, p_2, ... p_n, p_i \in \mathbb{N^{1,n}}$ \\
        \For{$i \gets 1$ \textbf{to} $n$}{
            $e_i \gets (p_i, v)$ \\
            $M \gets M \cup \{e_i\}$
        }
        \caption{Online AP with Advice}
    \end{algorithm}
    We define $A$ in our algorithm as an $n \times n$ matrix that corresponds to the bipartite graph with $a_{ij}$ being the weight from node $i$ of set 1 vertices to node $j$ of set 2 vertices. 
    
    The algorithm then, assumes that a certain ML model can predict a matrix $A'$ before the online assignment begin which is good enough to be used as the actual matching for the online assignment. It uses an offline algorithm (\cite{karp1980algorithm} in this instance) as an offline matching algorithm to produce optimal solutions from the prediction. This is the pre-calculation stage of the algorithm where we construct the data which the offline algorithm will use as a \textit{lookup} to make better predictions. 
    
    With the obtained $A'$, we can compute the matching $P$ which is a matrix that can be projected to A, such that we get the proposed optimal solution for $A$ with respect to the pre-processed $A'$. 
    
    In a certain sense, this separates the Machine Learning Model into our algorithm. It makes sense that for this study, we treat the ML Model as a black box that produces a prediction matrix $A'$ subject to some error $E$. With this, the study will revolve around empirical tests of these parameters and analyzing how they affect the solution quality of the online algorithm. The figure below shows the relationship and the parameters that will be used in this paper to investigate the effectiveness of this algorithm with the Online Assignment Problem.
    \begin{figure}[h]
      \centering
      \includegraphics[width=\linewidth]{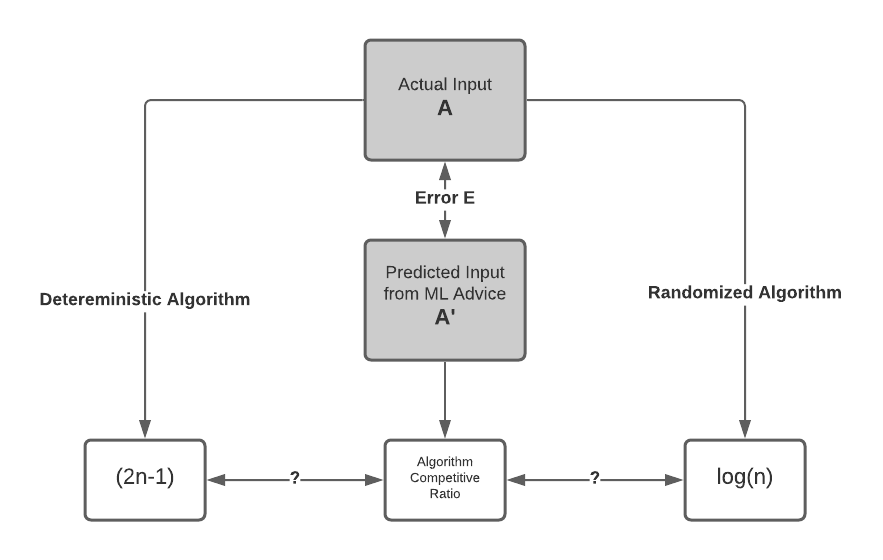}
      \caption{Framework on the process of investigating the proposed Algorithm for the Online Assignment Problem}
        \Description{Diagram.}
    \end{figure}
    
    The diagram above shows the relationship between both actual inputs $A$ and predicted input $A'$, we introduce an error parameter $E$ that describes the total distance between $A$ and $A'$. $E$ will be vital in the empirical testing for this algorithm, as different solution qualities may be produced for different values for $E$. 
    
    The algorithm then proceeds to using these matrices to perform online (for $A$) and offline (for $A'$) computations. The actual input $A$ will be computed using the best online algorithms for both deterministic (Khuller, 1994 \cite{khuller1994line}) and randomized (Bansal, 2007 \cite{bansal2007log}). An offline algorithm will then be used to get the solution for the predicted input $A'$ and all the solutions will be compared which will be the main point of analysis for this study. 
    
    Our experimentation will be using the python library \textit{networkx} and will use the Khuller, 1994 offline matching algorithm in getting the matching for the predicted matrix $A'$ which will be projected into $A$.
    
    \subsection{Running Time Analysis}
    Theoretically lines 5-7 of the algorithm runs at $O(n)$ time as it performs lookup using advice, while line 4 executes an offline algorithm computation for the perturbed matrix and runs at $O(n^2\log{n})$ \cite{karp1980algorithm}. Lines 1-3 are the portions of the algorithm which uses a machine learning model that is treated as a black box, thus, the running time for this depends on how testing with a machine learning models is executed. As such, it is widely accepted that most ML Models provide better predictions with more training time which should also be considered when analyzing the total running time for this algorithm.
    
    For the purposes of this study, the table below shows a bar graph of the empirical running time obtained by the implementation of the algorithm using Python and NetworkX
    
    \begin{figure}[h]
      \centering
      \includegraphics[width=\linewidth]{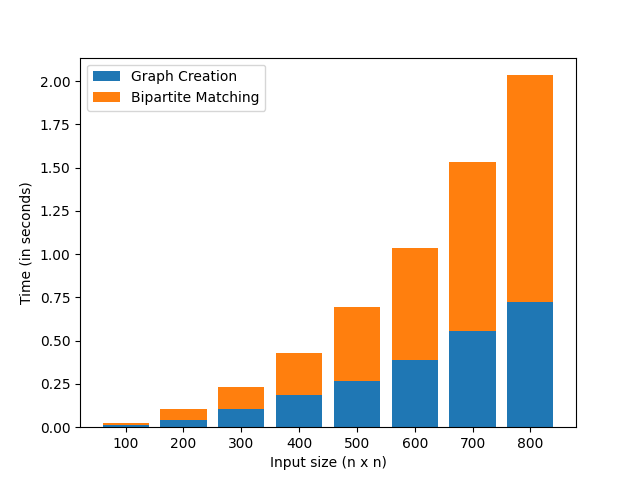}
      \caption{A bar graph of the empirical time of implementation in Python and NetworkX}
        \Description{Diagram.}
    \end{figure}
    
    
    \subsection{Parameter Definitions}
    In this empirical testing, we first define A. In our experimentation, we used an unbiased test data set collected from Beasley, 1990 \cite{beasley1990linear} as our input data on which we would test our algorithm using different error parameters on perturbing and differentiating $E$. The Beasley data set is taken from a collection of benchmark test data sets for Operational Research and is produced with Linear Programming with Cray Supercomputers. 
    
    The Beasley assignment data set which we will denote as $A_{[B]}$ can be defined as a matrix:
    \begin{center}
        $n \times n$ where for all $a_{[B]} \in A_{[B]} $, $a_{ij[B]} \in \{1,2,...100\}$
    \end{center} 
    \noindent \textbf{Error for Beasley Matrices.} We define for this study different methods on introducing errors to identify which parameters of the matrix affect the resulting competitive ratio for this algorithm. For all Beasley matrices, we denote $E_B$ as the total error (distance) of $A_{B}$ to $A^{'}_{B}$ defined as
    \[E_B = \epsilon n^2 \cdot \textsc{Incr}(\mu)\]
    Where we define $\epsilon$ as a parameter for empirical testing which controls the frequency of elements of $A_{[B]}$ to be perturbed, $n^2$ as the size of the matrix, $\textsc{Incr}$ as the method of getting the numerical value that will be perturb to an element. It uses a parameter $\mu$ to control the size for empirical testing. 
    
    We also define $\textsc{Rand}(A_{[B]}|\epsilon)$ as a method of selecting elements of the matrix to be perturbed, this method uses the \textit{python library: Rand} to have uniformly distributed choices on elements to perturb. The output of this method is a matrix the same size as $A_{[B]}$ with uniformly distributed $1s$ on perturbed elements and 0 otherwise. 
    
    Using these, we can define the predicted matrix $A_{[B]}'$ as follows. 
    \[
        A_{[B]}' = A{[B]} + (\textsc{Rand}(A_{[B]}|\epsilon) \cdot \textsc{Incr}(\mu))
    \]
    
    In the following sections, we will define 2 variations \textsc{Incr} for the Beasley data set which this study will use in its empirical testing and analysis. 
    
    \noindent \textbf{Error with respect to max value.} We define \textsc{Incr} as scaling the value of the maximum element of $A$ by $\mu$. 
    \[
    \textsc{Incr}(\mu) =
    \begin{cases}
    -\mu \cdot \textsc{max}(A_{[B]}); & a_{ij} + (\mu \cdot \textsc{max}(A_{[B]})) > \textsc{max}(A_{[B]}) \\
    \mu \cdot \textsc{max}(A_{[B]}); & a_{ij} + (\mu \cdot \textsc{max}(A_{[B]})) \leq \textsc{min}(A_{[B]}) \\
    \mu \cdot \textsc{max}(A_{[B]}); & \textsc{Python Rand} \geq 0.5\\
    -\mu \cdot \textsc{max}(A_{[B]}); & \textsc{Python Rand} < 0.5
    \end{cases}
    \]
    The piece wise definition of \textsc{Incr} ensures that the perturbation of the matrix still lies inside the bounds of the values of the original matrix. We denote the total error following this method as $E_\alpha$
    \[
    E = \epsilon n^2 \cdot|\mu\cdot max(A_{[B]})|
    \]
    By using this kind of perturbation to the actual matrix, we get to analyze how scaling the individual perturbation value with the maximum element of the matrix. With this, we get a static value to increment to the entire matrix that adjusts to the matrix. Later in the paper, we analyze how changing the values of the parameters for this method affects the solution quality of the algorithm. 

    \section{Results}
        \textbf{An investigation of the Algorithm}. Using the Beasley data set, we used the parameter definitions above to measure how the algorithm perform against the optimal deterministic and randomized algorithm. We will be showing results of our testing both with varying graph sizes, $\mu$ and $\epsilon$. We used the following values from table 3 as the benchmark values of our testing. 
        
        \begin{table}[h]
          \caption{Chosen benchmark values for empirical testing}
          \label{tab:freq}
          \begin{tabular}{ccl}
            \toprule
            Benchmark&Values\\
            \midrule
            Graph Size ($n$)& $[100, 200, 300, 400, 500, 600, 700, 800]$\\
            $\epsilon$  & $[0, 0.1, 0.2, 0.3, 0.4, 0.5]$\\
            $\mu$ & $[0.1, 0.3, 0.5]$\\
          \bottomrule
        \end{tabular}
        \end{table}
    
    The choice of values for $\mu$ and $\epsilon$ is derived from the possible expected error incurred from a Machine Learning technique and is incremented to identify its impact on the resulting solution quality values. Limiting the error metrics to $50\%$ is a design choice for the empirical analysis of this study as we believed that the advancement of ML Models are advanced enough to not incur more than $50\%$ for both $\mu$ and $\epsilon$.
    
    The values presented below, are selected resulting competitive ratios with the given benchmarks and varying values of $\mu$ scaling with the max value. 
    
    \begin{table}
        \centering
        \sisetup{round-mode=places,round-precision=1}
        \caption{Resulting competitive ratios using presented algorithm with $\mu = 0.1, 0.2, 0.3$}
        \begin{tabular}{@{} l *{8}{S[table-format=-1.2]} @{}} 
        \toprule
        $\epsilon \backslash n$& {100} & {200} & {300} & {400} & {500}& {600} & {700} & {800}\\
        \midrule
        0    & 1.0 & 1.0& 1.0 & 1.0& 1.0 & 1.0& 1.0 & 1.0\\
        0.1   &  1.3442622950819672& 1.6652631578947368& 2.1054313099041533& 2.4054726368159205& 2.547931382441978& 2.7908163265306123& 3.066079295154185& 2.9594072164948453\\
        0.2  &  1.6688524590163933& 2.134736842105263& 2.840255591054313&3.054726368159204& 3.4137235116044398&3.6607142857142856&3.969897209985316&4.010953608247423\\
        0.3     &  1.9180327868852458& 2.6757894736842105& 3.36741214057508& 3.531094527363184& 4.0020181634712415&4.2270408163265305&4.451541850220265& 4.501288659793815\\
        0.4 & 2.2295081967213113& 2.9894736842105263&3.916932907348243& 4.181592039800995& 4.372351160443996&4.624149659863946&4.7804698972099855&4.826675257731959\\
        0.5 & 2.678688524590164&3.263157894736842&4.142172523961661& 4.407960199004975& 4.580221997981837&4.877551020408164&5.046989720998532&5.038015463917525\\
        \bottomrule
        \end{tabular}
        \begin{tabular}{@{} l *{8}{S[table-format=-1.2]} @{}} 
        \toprule
        $\epsilon \backslash n$& {100} & {200} & {300} & {400} & {500}& {600} & {700} & {800}\\
        \midrule
        0    & 1.0 & 1.0& 1.0 & 1.0& 1.0 & 1.0& 1.0 & 1.0\\
        0.1   &  2.3049180327868855& 3.383157894736842& 5.110223642172524& 6.0& 6.769929364278506& 7.082482993197279& 7.676945668135096& 8.072809278350515\\
        0.2  &  3.1508196721311474&4.7642105263157895&6.862619808306709& 8.449004975124378&  9.047426841574168& 10.33078231292517& 10.411160058737151& 10.999355670103093\\
        0.3     &  4.239344262295082& 6.551578947368421& 8.777955271565496& 10.17039800995025& 10.643794147325933&11.988945578231293& 12.387665198237885&13.100515463917526\\
        0.4 & 4.947540983606557& 7.44&9.768370607028753&11.106965174129353& 12.360242179616549&13.118197278911564&13.734214390602055& 14.13853092783505\\
        0.5 & 6.344262295081967&8.650526315789474&10.936102236421725&12.573383084577115& 13.143289606458124&14.0671768707483&14.68575624082232& 14.938144329896907\\
        \bottomrule
        \end{tabular}
        \begin{tabular}{@{} l *{8}{S[table-format=-1.2]} @{}} 
        \toprule
        $\epsilon \backslash n$& {100} & {200} & {300} & {400} & {500}& {600} & {700} & {800}\\
        \midrule
        0    & 1.0 & 1.0& 1.0 & 1.0& 1.0 & 1.0& 1.0 & 1.0\\
        0.1   &  3.6885245901639343& 6.557894736842106& 8.578274760383387& 10.82960199004975& 12.807265388496468& 14.439625850340136& 15.447870778267253&16.22873711340206\\
        0.2  &  5.9573770491803275&11.983157894736841& 14.207667731629392& 16.34328358208955& 17.25832492431887&19.869897959183675& 20.701174743024964&22.010953608247423\\
        0.3     &  8.20983606557377&14.336842105263157& 16.731629392971247&19.83830845771144& 21.5146316851665&22.33078231292517& 23.270190895741557&24.409149484536083\\
        0.4 & 10.770491803278688& 15.75578947368421& 19.522364217252395& 22.036069651741293& 23.489404641775984& 24.128401360544217& 24.54625550660793& 25.36791237113402\\
        0.5 & 11.839344262295082& 17.370526315789473& 21.121405750798722& 23.833333333333332& 24.322906155398588& 24.877551020408163& 25.394273127753305&25.736469072164947\\
        \bottomrule
        \end{tabular}
    \end{table}
    \begin{figure}[h]
      \centering
      \includegraphics[scale=0.5]{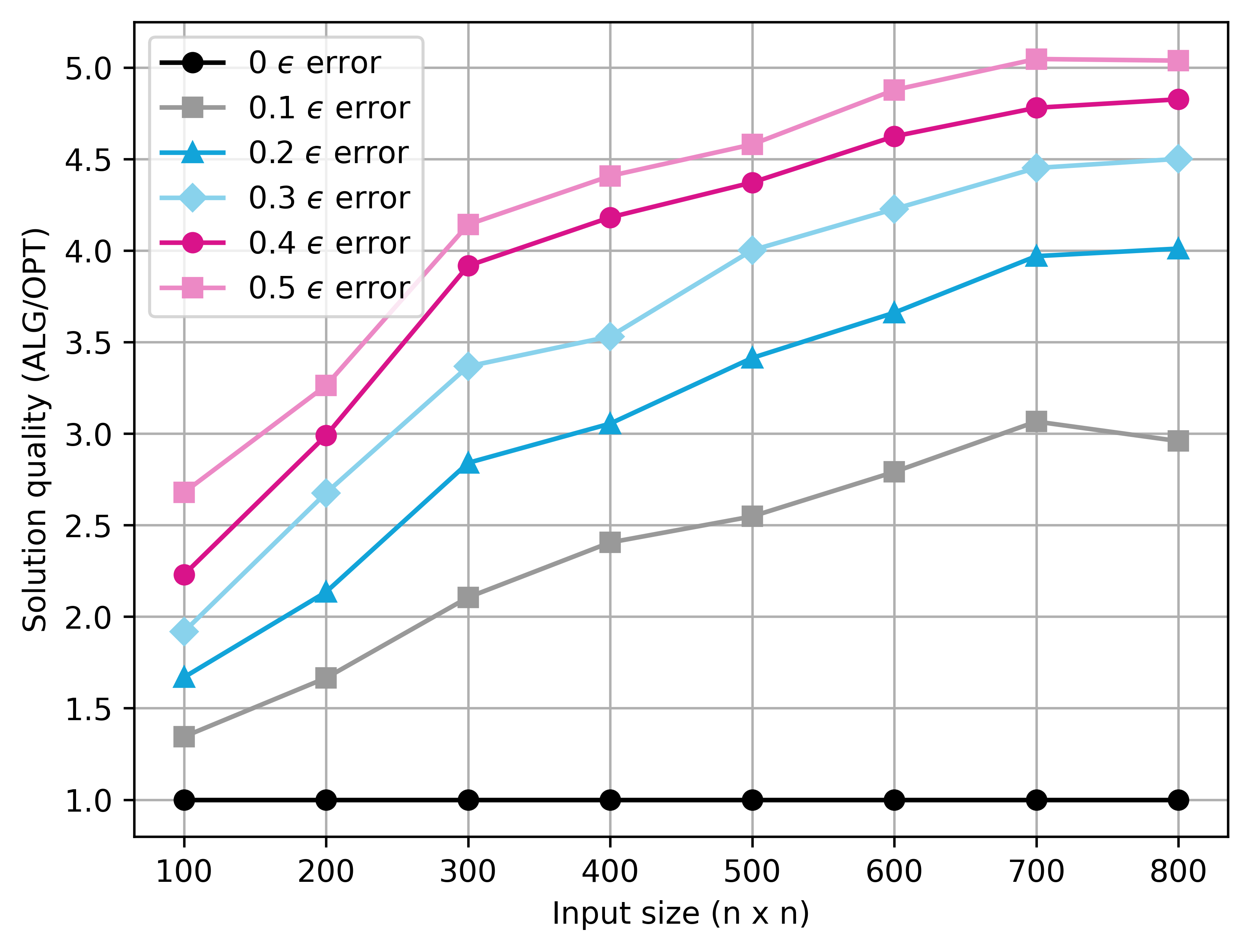}
      \caption{Error Graph using $A_{[B]}$ and Matching obtained from $A_{[B]}'$ with $\mu=0.1$}
        \Description{Diagram.}
    \end{figure}
    \begin{figure}[h]
      \centering
      \includegraphics[scale=0.5]{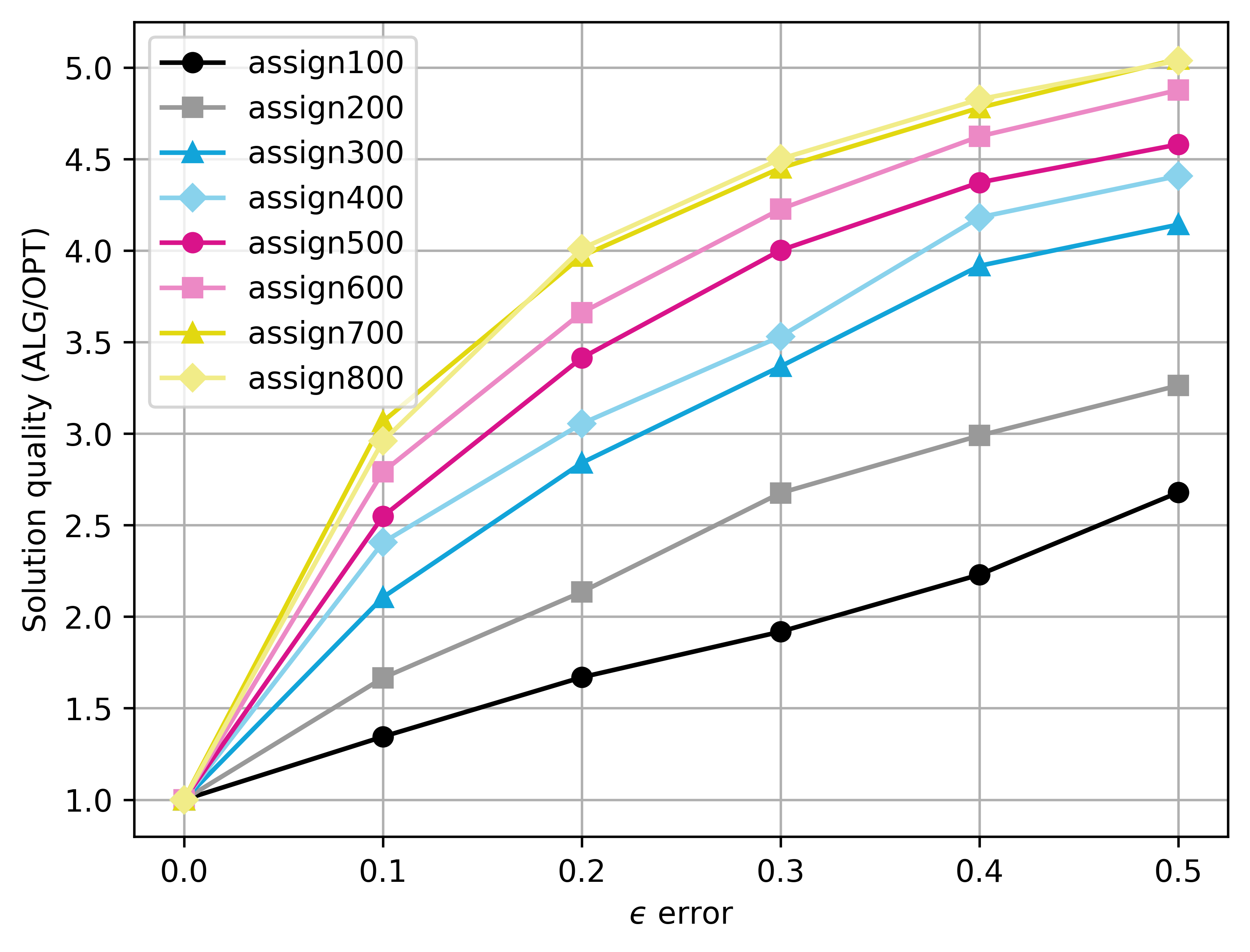}
      \caption{Size Graph using $A_{[B]}$ and Matching obtained from $A_{[B]}'$ with $\mu=0.1$}
        \Description{Diagram.}
    \end{figure}
    \begin{figure}[h]
      \centering
      \includegraphics[scale=0.5]{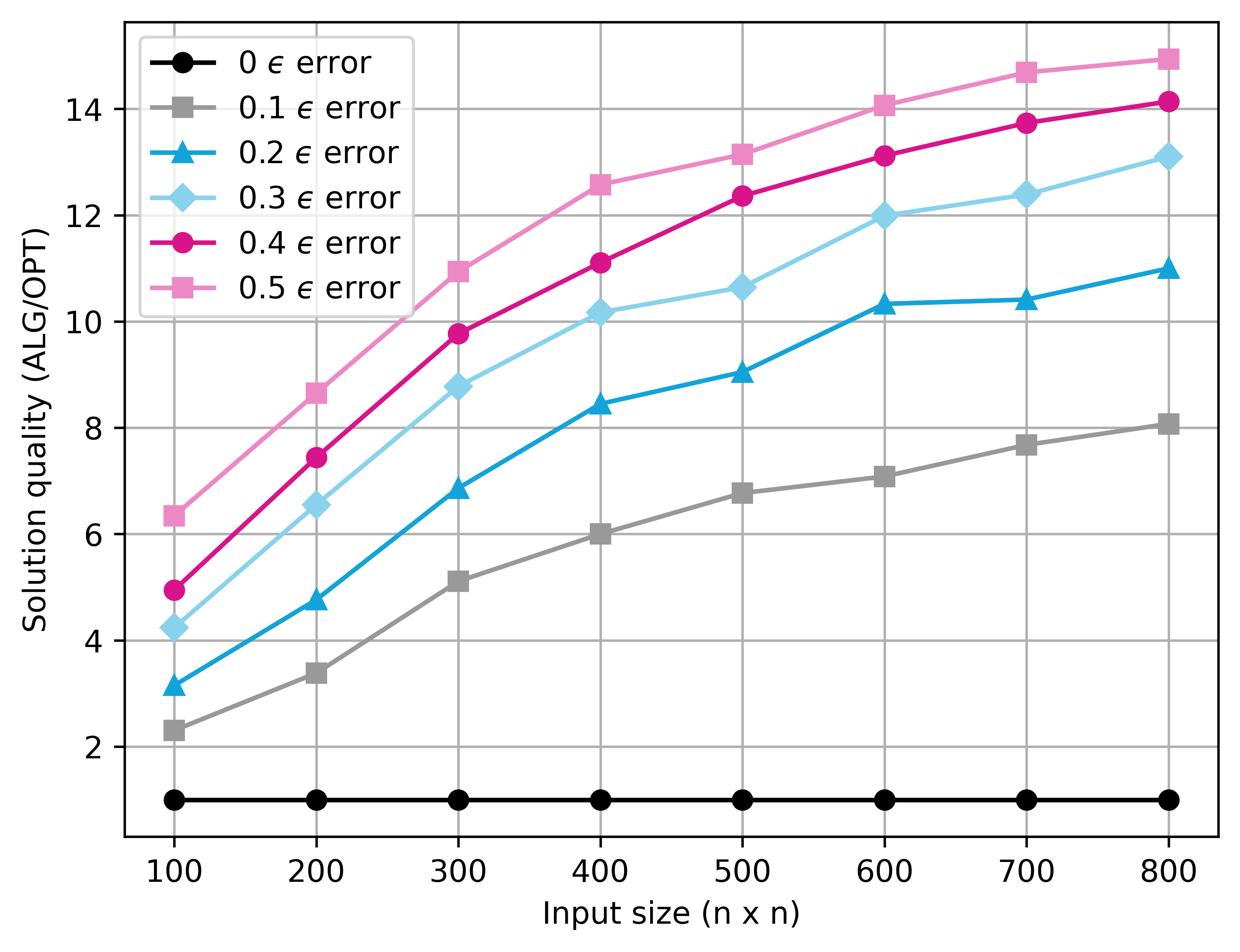}
      \caption{Error Graph using $A_{[B]}$ and Matching obtained from $A_{[B]}'$ with $\mu=0.3$}
        \Description{Diagram.}
    \end{figure}
    \begin{figure}[h]
      \centering
      \includegraphics[scale=0.5]{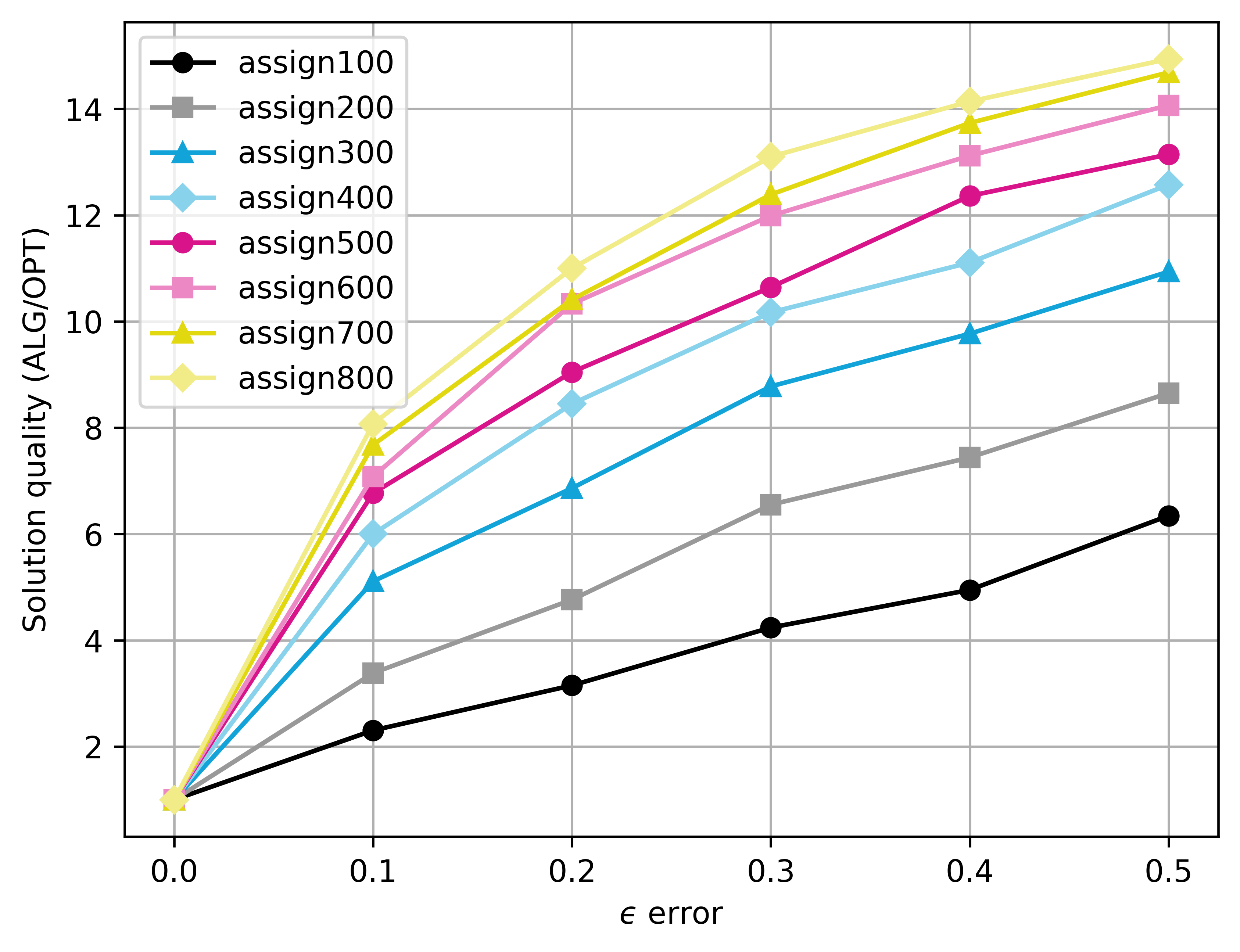}
      \caption{Size Graph using $A_{[B]}$ and Matching obtained from $A_{[B]}'$ with $\mu=0.3$}
        \Description{Diagram.}
    \end{figure}
    \begin{figure}[h]
      \centering
      \includegraphics[scale=0.5]{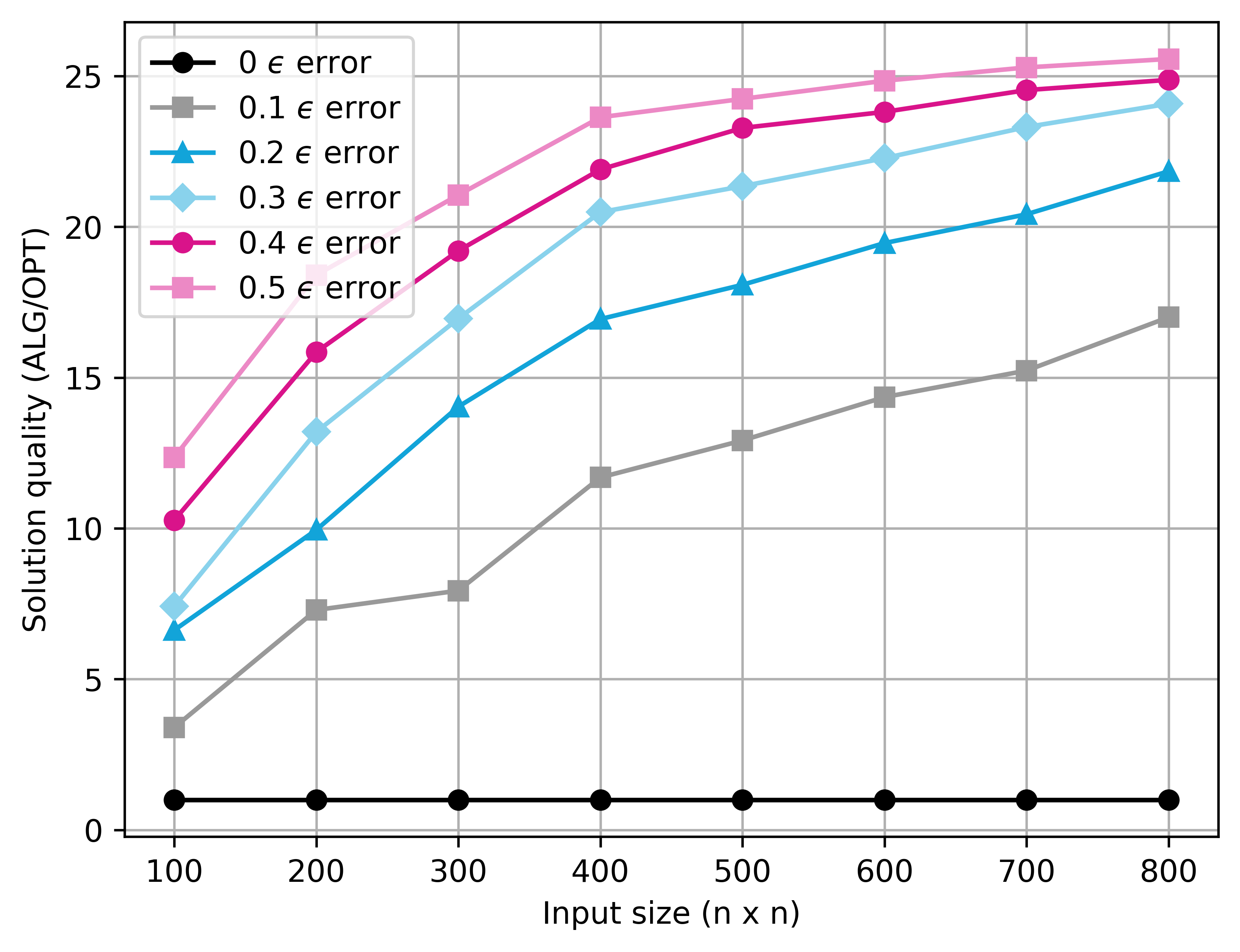}
      \caption{Error Graph using $A_{[B]}$ and Matching obtained from $A_{[B]}'$ with $\mu=0.5$}
        \Description{Diagram.}
    \end{figure}
    \begin{figure}[h]
      \centering
      \includegraphics[scale=0.5]{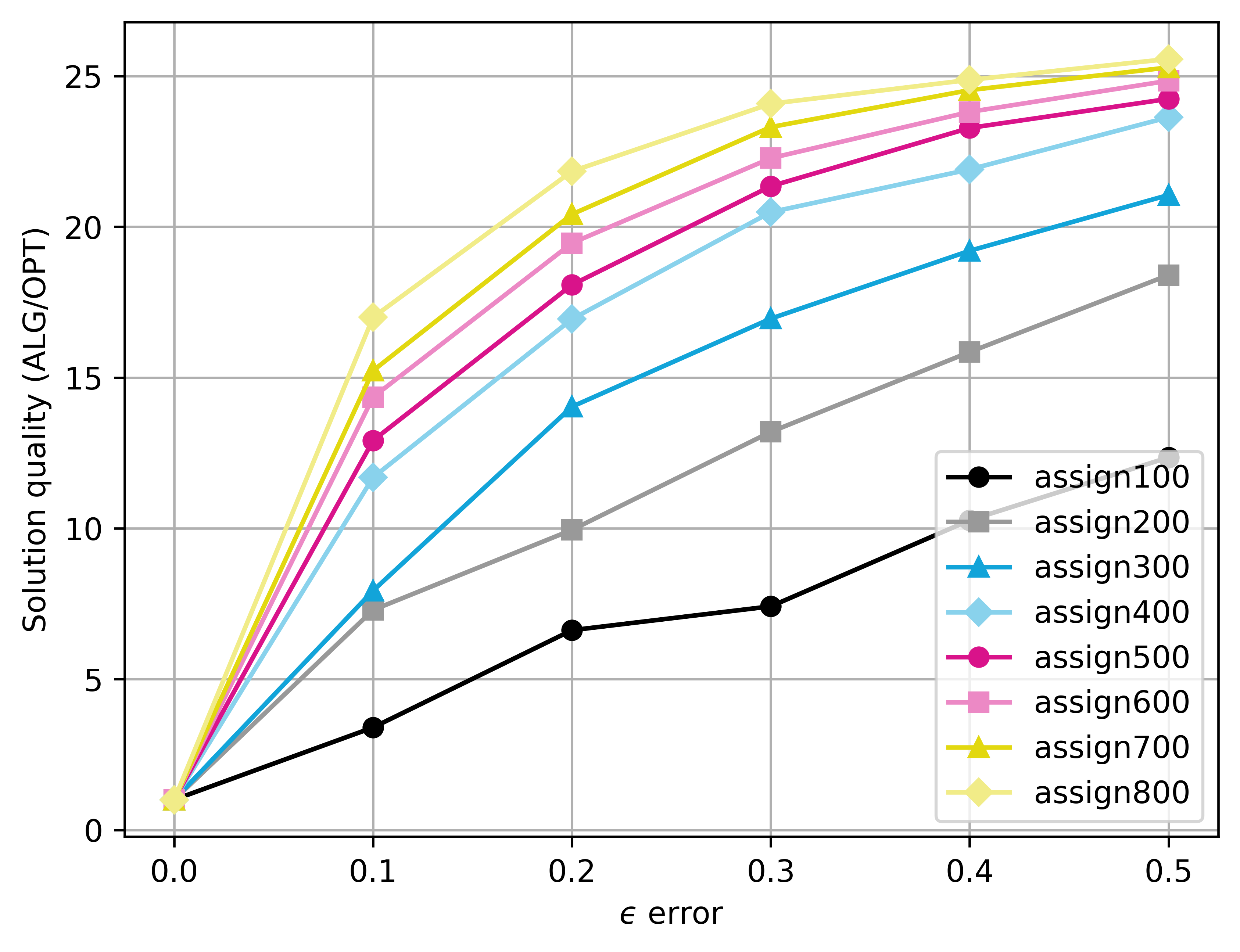}
      \caption{Size Graph using $A_{[B]}$ and Matching obtained from $A_{[B]}'$ with $\mu=0.5$}
        \Description{Diagram.}
    \end{figure}

    Though the values vary randomly, which is expected as we gained resulting solution from the projecting the matching obtained from the predicted graph $A_{[B]}$ into $A_{[B]}'$, it can be observed the direct proportionality between the competitive ratio with the size and error. We can infer with this discovery, that for the Beasley assignment data set, we get a better competitive ratio for decreasing error or the closer the $A_{[B]}'$ with $A_{[B]}$, the solution quality increases. 

 \textbf{Benchmark Data Set.} As this study included a process of perturbing a known benchmark data set, further studies with the algorithm and can use the perturbed data as input for other online algorithms with untrusted advice. The definition of $\mu$ and $\epsilon$ together with the discoveries in the analysis section of this paper can also be used as a reference upon generation of input sequence as untrusted advice. 
    
    \subsection{Analysis}
    As it was quite easy to see that with increasing $n$, the competitive ratio increases, which is also consistent with the known deterministic and randomized algorithms. Though, we can infer from the results that the size of the matrix is not the main driving force in getting a worse competitive ratio. The $\mu$ on the other hand had a more drastic effect on the competitive ratio of the algorithm. For the Beasley data set, the max value of all matrices are 100, which means that the values added or subtracted using the single element perturbation from $\textsc{Incr}$ are 10, 30 and 50. Increasing the value of $\mu$ for a predicted matrix with $\epsilon = 0.1$ and size of 800, gives a competitive ratio of 5, 14.9 and 25.7 respectively which is a steeper change in competitive ratio. This gives us the conclusion that the size of change of perturbed elements has the greatest effect on the competitive ratio for the algorithm for the Beasley data set. 
    
    We now try to discover how well our algorithm perform against the best randomized and deterministic algorithms. It is apparent that for the Beasley Data set, our algorithm performs much better than the tight bounded \cite{khuller1994line} Khuller, 1994 algorithm that has a competitive ratio of $2n-1$. We therefore proceed to comparing the competitive ratio with the best randomized algorithm from \cite{bansal2007log} of $\log^2 (n)$ and the tight bound proven as $\ln(k)$. For the Beasley Data set with max scaled perturbation, our testing showed that this algorithm performs better for all benchmark sizes $n$ and $\epsilon$ when $\mu = 0.1$
    \begin{figure}[h]
      \centering
      \includegraphics[scale=0.5]{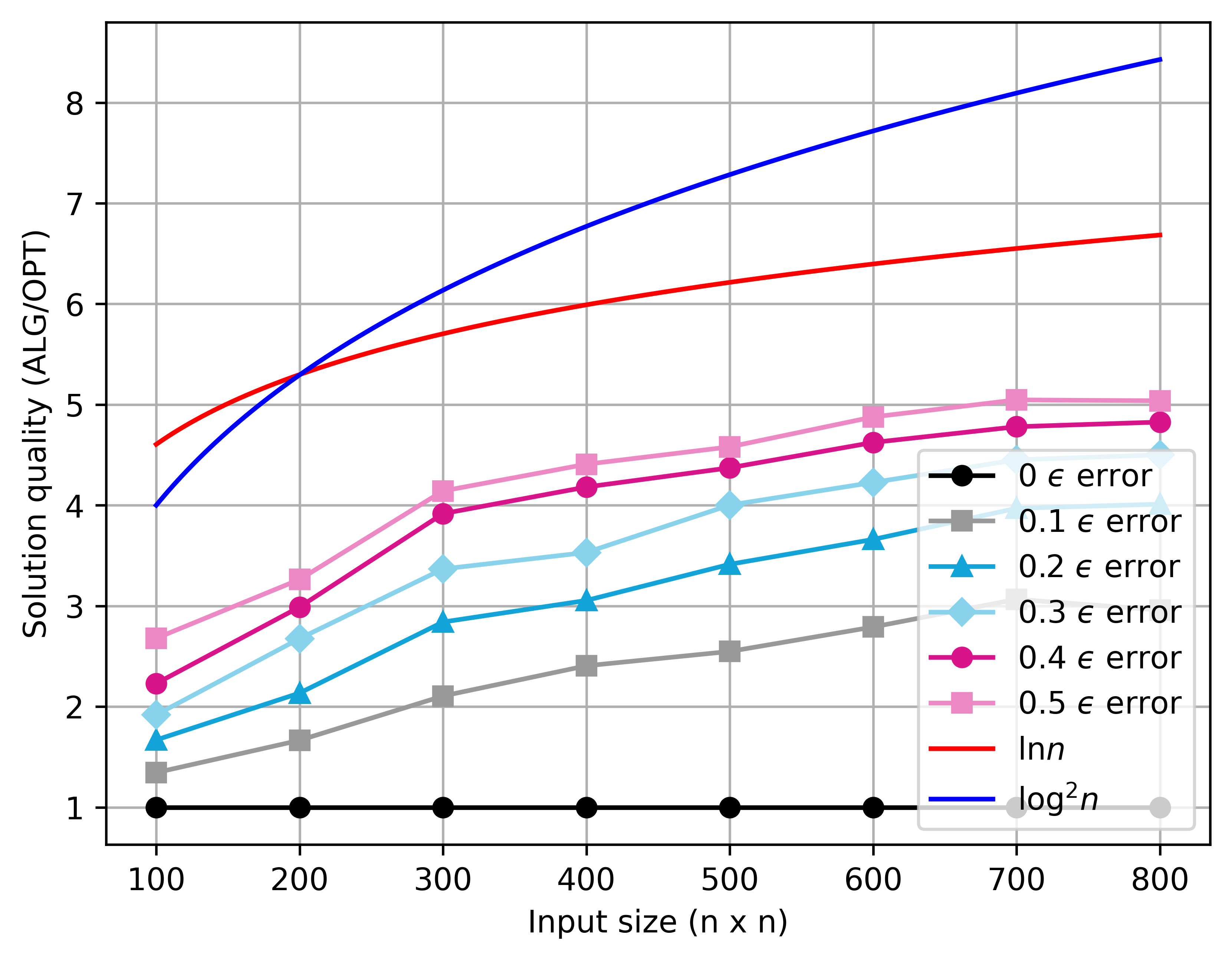}
      \caption{Graphical comparison between competitive ratios of the presented algorithm and best known randomized algorithms with $\mu=0.1$}
        \Description{Diagram.}
    \end{figure}
    
    The algorithm also performs relatively well with $\mu=0.3$ on when $\epsilon = 0.1$ or below, specifically, our algorithms beat $\log^2(n)$ on all sizes, but fails to beat the competitive ratio of $\ln(k)$ for Beasley matrices of $n = 500$ and above, but performs better when $\epsilon=0.2$ on matrices of $size = 200$ and below for both randomized algorithms. This tells us that all of the directly proportional parameter values  contribute into having a better solution for this algorithm. Thus, with a sufficient enough Machine Learning algorithm that can give predictions under these threshold, our algorithm will be able to perform relatively better against the best online deterministic and randomized algorithms on their resulting competitive ratios. 
    
    \section{Future Work}
    As this study only focuses on empirical analysis from the benchmark data set and our produced algorithm, a theoretical estimate of the closed form of the function obtained from the algorithm that describes the relationship between the error and solution quality with respect to $\mu$ and $\epsilon$ would be greatly valued. 
    
    With regards to the Machine Learning algorithm, the study treated ML as a black box in which ML is purely represented as a device that returns a perturbed prediction matrix to be used in computation. Studies whether the running time of running such ML techniques together with the described algorithm will be optimal, also studies on different techniques and how they affect the algorithm can be a focus on future work. While we investigated and analyzed the effectiveness of the presented algorithm, the ML model that will be used in obtaining the predicted advice matrix may be a limitation for the process especially with regards to the overall running time of the algorithm. 
    
    Also, the use of this algorithm with other numerical spaces can be further studied as the empirical tests of this paper only focused on known benchmark data set. This goes the same with different definitions of error metrics $\mu$ and $\epsilon$ which can help provide further analysis of the algorithm. 
    
    \begin{figure}[h]
      \centering
      \includegraphics[scale=0.5]{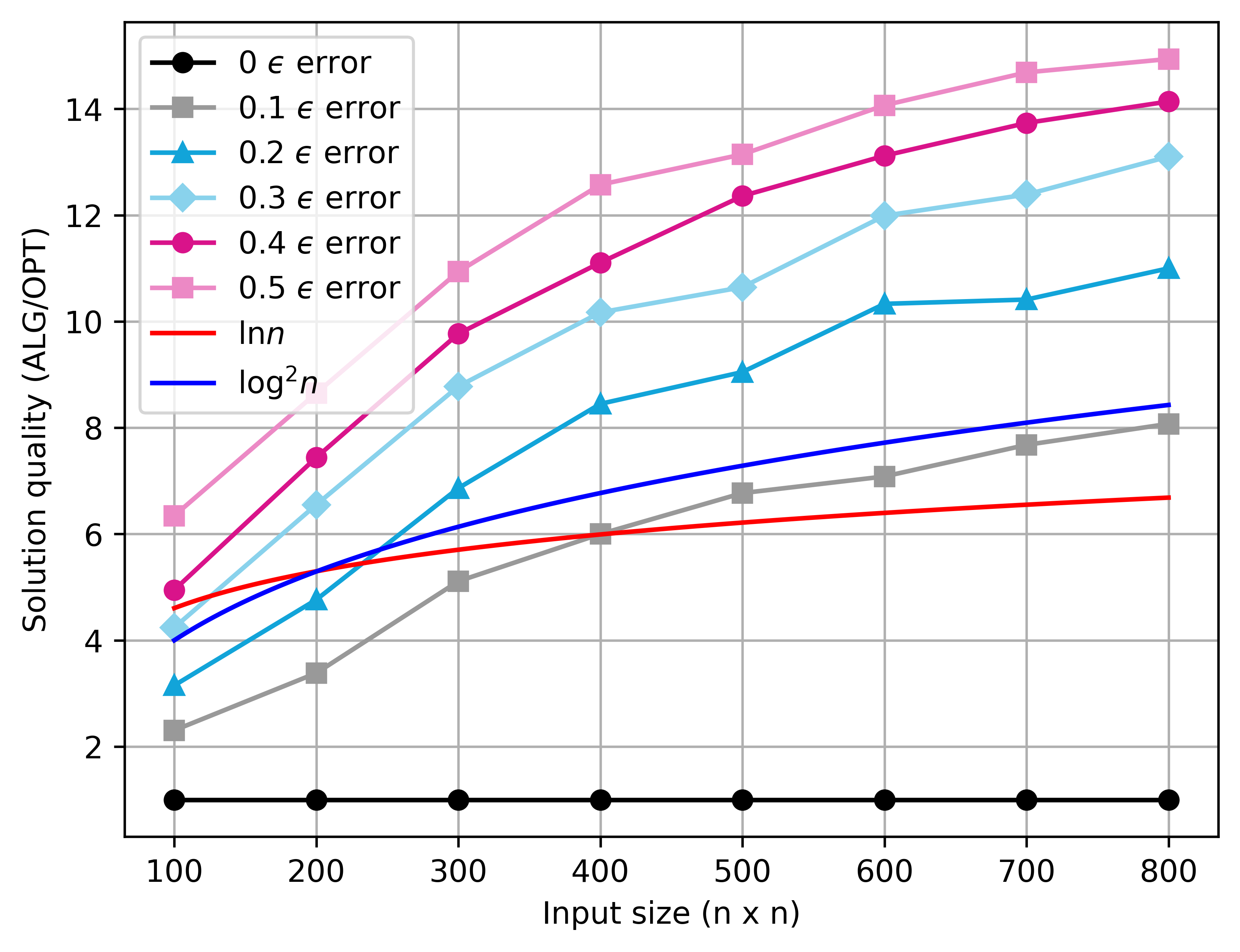}
      \caption{Graphical comparison between competitive ratios of the presented algorithm and best known randomized algorithms with $\mu=0.3$}
        \Description{Diagram.}
    \end{figure}



\bibliographystyle{ACM-Reference-Format}
\bibliography{main-bib}


\begin{thebibliography}{19}


\ifx \showCODEN    \undefined \def \showCODEN     #1{\unskip}     \fi
\ifx \showDOI      \undefined \def \showDOI       #1{#1}\fi
\ifx \showISBNx    \undefined \def \showISBNx     #1{\unskip}     \fi
\ifx \showISBNxiii \undefined \def \showISBNxiii  #1{\unskip}     \fi
\ifx \showISSN     \undefined \def \showISSN      #1{\unskip}     \fi
\ifx \showLCCN     \undefined \def \showLCCN      #1{\unskip}     \fi
\ifx \shownote     \undefined \def \shownote      #1{#1}          \fi
\ifx \showarticletitle \undefined \def \showarticletitle #1{#1}   \fi
\ifx \showURL      \undefined \def \showURL       {\relax}        \fi
\providecommand\bibfield[2]{#2}
\providecommand\bibinfo[2]{#2}
\providecommand\natexlab[1]{#1}
\providecommand\showeprint[2][]{arXiv:#2}

\bibitem[\protect\citeauthoryear{Bansal, Buchbinder, Gupta, and Naor}{Bansal
  et~al\mbox{.}}{2007}]%
        {bansal2007log}
\bibfield{author}{\bibinfo{person}{Nikhil Bansal}, \bibinfo{person}{Niv
  Buchbinder}, \bibinfo{person}{Anupam Gupta}, {and}
  \bibinfo{person}{Joseph~Seffi Naor}.} \bibinfo{year}{2007}\natexlab{}.
\newblock \showarticletitle{An O (log 2 k)-competitive algorithm for metric
  bipartite matching}. In \bibinfo{booktitle}{\emph{European symposium on
  algorithms}}. Springer, \bibinfo{pages}{522--533}.
\newblock


\bibitem[\protect\citeauthoryear{Beasley}{Beasley}{1990}]%
        {beasley1990linear}
\bibfield{author}{\bibinfo{person}{JE Beasley}.}
  \bibinfo{year}{1990}\natexlab{}.
\newblock \showarticletitle{Linear programming on Cray supercomputers}.
\newblock \bibinfo{journal}{\emph{Journal of the Operational Research Society}}
  \bibinfo{volume}{41}, \bibinfo{number}{2} (\bibinfo{year}{1990}),
  \bibinfo{pages}{133--139}.
\newblock


\bibitem[\protect\citeauthoryear{B{\"o}ckenhauer, Komm, Kr{\'a}lovi{\v{c}},
  Kr{\'a}lovi{\v{c}}, and M{\"o}mke}{B{\"o}ckenhauer et~al\mbox{.}}{2009}]%
        {bockenhauer2009advice}
\bibfield{author}{\bibinfo{person}{Hans-Joachim B{\"o}ckenhauer},
  \bibinfo{person}{Dennis Komm}, \bibinfo{person}{Rastislav
  Kr{\'a}lovi{\v{c}}}, \bibinfo{person}{Richard Kr{\'a}lovi{\v{c}}}, {and}
  \bibinfo{person}{Tobias M{\"o}mke}.} \bibinfo{year}{2009}\natexlab{}.
\newblock \showarticletitle{On the advice complexity of online problems}. In
  \bibinfo{booktitle}{\emph{International Symposium on Algorithms and
  Computation}}. Springer, \bibinfo{pages}{331--340}.
\newblock


\bibitem[\protect\citeauthoryear{Dinic and Kronrod}{Dinic and Kronrod}{1969}]%
        {dinic1969algorithm}
\bibfield{author}{\bibinfo{person}{EA Dinic} {and} \bibinfo{person}{MA
  Kronrod}.} \bibinfo{year}{1969}\natexlab{}.
\newblock \showarticletitle{An algorithm for the solution of the assignment
  problem}. In \bibinfo{booktitle}{\emph{Soviet Math. Dokl}},
  Vol.~\bibinfo{volume}{10}. \bibinfo{pages}{1324--1326}.
\newblock


\bibitem[\protect\citeauthoryear{Dobrev, Kr{\'a}lovi{\v{c}}, and
  Pardubsk{\'a}}{Dobrev et~al\mbox{.}}{2008}]%
        {dobrev2008much}
\bibfield{author}{\bibinfo{person}{Stefan Dobrev}, \bibinfo{person}{Rastislav
  Kr{\'a}lovi{\v{c}}}, {and} \bibinfo{person}{Dana Pardubsk{\'a}}.}
  \bibinfo{year}{2008}\natexlab{}.
\newblock \showarticletitle{How much information about the future is needed?}.
  In \bibinfo{booktitle}{\emph{International Conference on Current Trends in
  Theory and Practice of Computer Science}}. Springer,
  \bibinfo{pages}{247--258}.
\newblock


\bibitem[\protect\citeauthoryear{Edmonds and Karp}{Edmonds and Karp}{1972}]%
        {edmonds1972theoretical}
\bibfield{author}{\bibinfo{person}{Jack Edmonds} {and}
  \bibinfo{person}{Richard~M Karp}.} \bibinfo{year}{1972}\natexlab{}.
\newblock \showarticletitle{Theoretical improvements in algorithmic efficiency
  for network flow problems}.
\newblock \bibinfo{journal}{\emph{Journal of the ACM (JACM)}}
  \bibinfo{volume}{19}, \bibinfo{number}{2} (\bibinfo{year}{1972}),
  \bibinfo{pages}{248--264}.
\newblock


\bibitem[\protect\citeauthoryear{Indyk, Mallmann-Trenn, Mitrovi{\'c}, and
  Rubinfeld}{Indyk et~al\mbox{.}}{2020}]%
        {indyk2020online}
\bibfield{author}{\bibinfo{person}{Piotr Indyk}, \bibinfo{person}{Frederik
  Mallmann-Trenn}, \bibinfo{person}{Slobodan Mitrovi{\'c}}, {and}
  \bibinfo{person}{Ronitt Rubinfeld}.} \bibinfo{year}{2020}\natexlab{}.
\newblock \showarticletitle{Online Page Migration with ML Advice}.
\newblock \bibinfo{journal}{\emph{arXiv preprint arXiv:2006.05028}}
  (\bibinfo{year}{2020}).
\newblock


\bibitem[\protect\citeauthoryear{Kalyanasundaram and Pruhs}{Kalyanasundaram and
  Pruhs}{1993}]%
        {kalyanasundaram1993online}
\bibfield{author}{\bibinfo{person}{Bala Kalyanasundaram} {and}
  \bibinfo{person}{Kirk Pruhs}.} \bibinfo{year}{1993}\natexlab{}.
\newblock \showarticletitle{Online weighted matching}.
\newblock \bibinfo{journal}{\emph{Journal of Algorithms}} \bibinfo{volume}{14},
  \bibinfo{number}{3} (\bibinfo{year}{1993}), \bibinfo{pages}{478--488}.
\newblock


\bibitem[\protect\citeauthoryear{Karp}{Karp}{1980}]%
        {karp1980algorithm}
\bibfield{author}{\bibinfo{person}{Richard~M Karp}.}
  \bibinfo{year}{1980}\natexlab{}.
\newblock \showarticletitle{An algorithm to solve the m$\times$ n assignment
  problem in expected time O (mn log n)}.
\newblock \bibinfo{journal}{\emph{Networks}} \bibinfo{volume}{10},
  \bibinfo{number}{2} (\bibinfo{year}{1980}), \bibinfo{pages}{143--152}.
\newblock


\bibitem[\protect\citeauthoryear{Khuller, Mitchell, and Vazirani}{Khuller
  et~al\mbox{.}}{1994}]%
        {khuller1994line}
\bibfield{author}{\bibinfo{person}{Samir Khuller}, \bibinfo{person}{Stephen~G
  Mitchell}, {and} \bibinfo{person}{Vijay~V Vazirani}.}
  \bibinfo{year}{1994}\natexlab{}.
\newblock \showarticletitle{On-line algorithms for weighted bipartite matching
  and stable marriages}.
\newblock \bibinfo{journal}{\emph{Theoretical Computer Science}}
  \bibinfo{volume}{127}, \bibinfo{number}{2} (\bibinfo{year}{1994}),
  \bibinfo{pages}{255--267}.
\newblock


\bibitem[\protect\citeauthoryear{Kuhn}{Kuhn}{1955}]%
        {kuhn1955hungarian}
\bibfield{author}{\bibinfo{person}{Harold~W Kuhn}.}
  \bibinfo{year}{1955}\natexlab{}.
\newblock \showarticletitle{The Hungarian method for the assignment problem}.
\newblock \bibinfo{journal}{\emph{Naval research logistics quarterly}}
  \bibinfo{volume}{2}, \bibinfo{number}{1-2} (\bibinfo{year}{1955}),
  \bibinfo{pages}{83--97}.
\newblock


\bibitem[\protect\citeauthoryear{Lattanzi, Lavastida, Moseley, and
  Vassilvitskii}{Lattanzi et~al\mbox{.}}{2020}]%
        {lattanzi2020online}
\bibfield{author}{\bibinfo{person}{Silvio Lattanzi}, \bibinfo{person}{Thomas
  Lavastida}, \bibinfo{person}{Benjamin Moseley}, {and} \bibinfo{person}{Sergei
  Vassilvitskii}.} \bibinfo{year}{2020}\natexlab{}.
\newblock \showarticletitle{Online scheduling via learned weights}. In
  \bibinfo{booktitle}{\emph{Proceedings of the Fourteenth Annual ACM-SIAM
  Symposium on Discrete Algorithms}}. SIAM, \bibinfo{pages}{1859--1877}.
\newblock


\bibitem[\protect\citeauthoryear{Lykouris and Vassilvitskii}{Lykouris and
  Vassilvitskii}{2018}]%
        {lykouris2018competitive}
\bibfield{author}{\bibinfo{person}{Thodoris Lykouris} {and}
  \bibinfo{person}{Sergei Vassilvitskii}.} \bibinfo{year}{2018}\natexlab{}.
\newblock \showarticletitle{Competitive caching with machine learned advice}.
\newblock \bibinfo{journal}{\emph{arXiv preprint arXiv:1802.05399}}
  (\bibinfo{year}{2018}).
\newblock


\bibitem[\protect\citeauthoryear{Meyerson, Nanavati, and Poplawski}{Meyerson
  et~al\mbox{.}}{2006}]%
        {meyerson2006randomized}
\bibfield{author}{\bibinfo{person}{Adam Meyerson}, \bibinfo{person}{Akash
  Nanavati}, {and} \bibinfo{person}{Laura Poplawski}.}
  \bibinfo{year}{2006}\natexlab{}.
\newblock \showarticletitle{Randomized online algorithms for minimum metric
  bipartite matching}. In \bibinfo{booktitle}{\emph{Proceedings of the
  seventeenth annual ACM-SIAM symposium on Discrete algorithm}}.
  \bibinfo{pages}{954--959}.
\newblock


\bibitem[\protect\citeauthoryear{Munkres}{Munkres}{1957}]%
        {munkres1957algorithms}
\bibfield{author}{\bibinfo{person}{James Munkres}.}
  \bibinfo{year}{1957}\natexlab{}.
\newblock \showarticletitle{Algorithms for the assignment and transportation
  problems}.
\newblock \bibinfo{journal}{\emph{Journal of the society for industrial and
  applied mathematics}} \bibinfo{volume}{5}, \bibinfo{number}{1}
  (\bibinfo{year}{1957}), \bibinfo{pages}{32--38}.
\newblock


\bibitem[\protect\citeauthoryear{Purohit, Svitkina, and Kumar}{Purohit
  et~al\mbox{.}}{2018}]%
        {purohit2018improving}
\bibfield{author}{\bibinfo{person}{Manish Purohit}, \bibinfo{person}{Zoya
  Svitkina}, {and} \bibinfo{person}{Ravi Kumar}.}
  \bibinfo{year}{2018}\natexlab{}.
\newblock \showarticletitle{Improving online algorithms via ml predictions}. In
  \bibinfo{booktitle}{\emph{Advances in Neural Information Processing
  Systems}}. \bibinfo{pages}{9661--9670}.
\newblock


\bibitem[\protect\citeauthoryear{Rohatgi}{Rohatgi}{2020}]%
        {rohatgi2020near}
\bibfield{author}{\bibinfo{person}{Dhruv Rohatgi}.}
  \bibinfo{year}{2020}\natexlab{}.
\newblock \showarticletitle{Near-optimal bounds for online caching with machine
  learned advice}. In \bibinfo{booktitle}{\emph{Proceedings of the Fourteenth
  Annual ACM-SIAM Symposium on Discrete Algorithms}}. SIAM,
  \bibinfo{pages}{1834--1845}.
\newblock


\bibitem[\protect\citeauthoryear{Steffen}{Steffen}{2014}]%
        {steffen2014advice}
\bibfield{author}{\bibinfo{person}{Bj{\"o}rn~C Steffen}.}
  \bibinfo{year}{2014}\natexlab{}.
\newblock \bibinfo{booktitle}{\emph{Advice complexity of online graph
  problems}}.
\newblock \bibinfo{publisher}{ETH Zurich}.
\newblock


\bibitem[\protect\citeauthoryear{Tomizawa}{Tomizawa}{1971}]%
        {tomizawa1971some}
\bibfield{author}{\bibinfo{person}{Nobuaki Tomizawa}.}
  \bibinfo{year}{1971}\natexlab{}.
\newblock \showarticletitle{On some techniques useful for solution of
  transportation network problems}.
\newblock \bibinfo{journal}{\emph{Networks}} \bibinfo{volume}{1},
  \bibinfo{number}{2} (\bibinfo{year}{1971}), \bibinfo{pages}{173--194}.
\newblock


\end{thebibliography}


\end{document}